\documentclass[twocolumn]{IEEEtran}

\usepackage{color}
\usepackage{amsmath}
\usepackage{amssymb}
\usepackage{graphicx}

\begin{document}

\title{
Belief-Propagation Decoding of Lattices \\ Using Gaussian Mixtures
}

\author{Brian~Kurkoski,~\IEEEmembership{Member,~IEEE,} Justin Dauwels,~\IEEEmembership{Member,~IEEE}%
\thanks{Submitted to \emph{Transactions on Information Theory} on October 18, 2008. }
\thanks{B. Kurkoski is with the University of Electro-Communications, Tokyo, Japan.   J. Dauwels with with Massachusetts Institute of Technology, Cambridge, MA.}
\thanks{B. Kurkoski was supported in part by the Ministry of Education, Science, Sports and Culture; Grant-in-Aid for Scientific Research (C) number 19560371. J.~Dauwels was supported in part by post-doctoral fellowships from the King Baudouin Foundation and the Belgian American Educational Foundation (BAEF). Part of this work was carried out at the RIKEN Brain Science Institute, Saitama, Japan.}%
\thanks{This work was presented in part at IEEE ISIT 2008, Toronto, Canada.} }

%\author{
%\authorblockN{Brian Kurkoski}
%\authorblockA{%Dept. of Information and Communications Engineering\\
%University of Electro-Communications\\
%Tokyo, Japan\\
%kurkoski@ice.uec.ac.jp}\\
%\and
%\authorblockN{Justin Dauwels}
%\authorblockA{
%Massachusetts Institute of Technology, Cambridge, MA\\
%Harvard Medical School, Boston, MA \\
%justin@dauwels.com}%
%}

%\renewcommand{\thefootnote}{\fnsymbol{footnote}}	
\renewcommand{\thefootnote}{\hfill}

\newcommand{\eqdef}{\stackrel{\scriptscriptstyle\bigtriangleup}{=} }
\newcommand{\Nbb}[3]{\mathcal{N}\!\left({#1};{#2},{#3}\right)}
\newtheorem{lemma}{Lemma}

\maketitle

 %\footnotetext{This research was partially supported by Grant-in-Aid for
%Scientific Research (C) number 1656XXXX, 2007.}

\begin{abstract}
A belief-propagation decoder for low-density lattice codes is given which represents messages explicitly as a mixture of Gaussians functions.  The key component is an algorithm for approximating a mixture of several Gaussians with another mixture with a smaller number of Gaussians.   This  Gaussian mixture reduction algorithm iteratively reduces the number of Gaussians by minimizing the distance between the original mixture and an approximation with one fewer Gaussians.  

Error rates and noise thresholds of this decoder are compared with those for the previously-proposed decoder which discretely quantizes the messages.   The error rates are indistinguishable for dimension 1000 and 10000 lattices, and the Gaussian-mixture decoder has a 0.2 dB loss for dimension 100 lattices.  The Gaussian-mixture decoder has a loss of about 0.03 dB in the noise threshold, which is evaluated via Monte Carlo density evolution. Further, the Gaussian-mixture decoder uses far less storage for the messages.
\end{abstract}

\section{Introduction}

Low-density lattice codes (LDLC) are lattices characterized by a sparse inverse generator matrix, making them suitable for  belief-propagation decoding.  Sommer, Feder and Shalvi proposed this lattice construction, described its belief-propagation  decoder, and gave extensive convergence analysis.  With decoding complexity which is linear in the dimension, it is possible to decode LDLC lattices with dimension of~$10^6$.    When used on the unconstrained-power AWGN channel, a noise threshold appeared within 0.6 dB of an asymptotic limit \cite{Sommer-it08}.  

A notable aspect of belief-propagation decoding of LDLC codes on AWGN channels is that the messages are mixtures of Gaussian functions.  This is appealing for a decoder implementation, but the number of Gaussians in the mixture grows doubly exponentially in the number of iterations, and a naive implementation using Gaussians is infeasible.  Thus, in prior work by Sommer et~al., the decoder messages were quantized.

This short paper describes a belief-propagation decoder for LDLC lattices which uses messages which are Gaussian mixture distributions.  That is, each message consists of means, variances and mixing coefficients.   The key part of this decoder is a Gaussian mixture reduction algorithm which approximates a Gaussian mixture distribution with several Gaussian components by another  mixture with fewer such components.   This algorithm stops the doubly exponential growth in the number of Gaussians.

This Gaussian mixture reduction algorithm compares the distance between all possible pairs of Gaussians in an input list, and replaces the closest pair with a single Gaussian.  This proceeds iteratively until a stopping condition is reached.  While the Kullback-Leibler (KL) divergence would be an appropriate distance measure, it does not have a closed form for Gaussian mixtures, and therefore it is not suitable for use in an efficient decoding algorithm.   Instead, the squared distance is used, which can be computed in closed form.   The special case of replacing two Gaussians with a single Gaussian, and the square distanced, is first described in Section \ref{sec:twogauss}.   Then, the Gaussian mixture reduction algorithm, which is the general case with an arbitrary number of Gaussians, is described in Section \ref{sec:moregauss}.  

A related technique is the iterative pairwise replacement algorithm for reducing the order of a Gaussian mixture in kernel density estimation \cite{Scott-technometrics01}.  For compressed sensing, this algorithm was applied to find the solution to an underdetermined system of equations, when there is an a priori distribution on the unknowns \cite{Sarvotham-TR-Rice-2006}.   This inference algorithm has some similarities with LDLC decoding.   In parallel to our work, alternative methods to reduce the order of a Gaussian mixture have recently been proposed for use in statistical learning applications \cite{Goldberger-pami08}.

LDLC lattice construction and the unconstrained-power communication system is reviewed in Section \ref{sec:sys}.   Then, Section \ref{sec:ldlc} presents the LDLC decoder which represents messages using the means, variance and mixing coefficients of the component Gaussians.   Gaussian mixtures were used by Sommer et~al. to analyze decoder convergence; the main novelty here is to show how the Gaussian mixture reduction algorithm is incorporated into such a decoder.    

The error-rate performance of the Gaussian-mixture decoder is either indistinguishable from, or close to, that of the quantized-message decoder, as is numerically demonstrated in Section \ref{sec:finite}.  The error rates of the Gaussian-mixture decoder and the quantized-message decoder are indistinguishable for dimension 1000 and 10000 lattices, and there is a loss of about 0.2 dB for dimension 100 lattices.   Additionally, to place the error rates of LDLC lattices in context, comparisons are made with a universal lattice sphere bound.   Dimension 100 LDLC lattices have a substantial gap with this bound, but this gaps decreases as the dimension increases.

Noise thresholds for LDLC codes, found by Monte Carlo density evolution, are used to evaluate the complexity-performance tradeoff in Section \ref{sec:complexity}.   In the high-complexity case, the noise thresholds for the Gaussian-mixture decoder lose 0.03 dB with respect to the quantized-message decoder.   On the other hand, complexity can be substantially reduced while decreasing the loss to 0.1 dB.  In practice, the number of Gaussians in the mixtures was small, and so the Gaussian-mixture decoder requires far less storage than the quantized-message decoder.   Discussion is given in Section \ref{sec:discussion}.

The following notation is used.  A Gaussian distribution with mean $m$ and variance $v$ is denoted as:
\begin{eqnarray}
\Nbb z m v = \frac{1}{\sqrt{2 \pi v}} e^{ - \frac{ (z-m)^2}{2 v}}.
\end{eqnarray}
A message $f(z)$ is a mixture of $N$ Gaussians:
\begin{eqnarray}
f(z) = \sum_{i=1}^{N} c_i \, \Nbb{z}{m_i}{v_i}, \label{eqn:mixture}
\end{eqnarray}
where $c_i \geq 0$ are mixing coefficients with $\sum_{i=1}^{N} c_i = 1$.  An equivalent representation of $f(z)$ is a list of $N$ triples of means, variances and mixing coefficients, $\{t_1, \ldots, t_N\}$ $=$ $\{(m_1, v_1, c_1), \ldots, (m_{N}, v_{N}, c_{N} ) \}$, and these two representations will be used interchangeably.

\section{Gaussian Mixture Reduction Algorithm \label{sec:gmr} }

\subsection{Approximating Two Gaussians with One Gaussian \label{sec:twogauss}}

When approximating a true distribution $p(z)$, by an approximate distribution $q(z)$, a reasonable approach is to choose $q(z)$ in such a way that the KL divergence, $\mathrm{KL}(p || q)$, is minimized.   For distributions with support $\mathcal Z$, the KL divergence is given by:
\begin{eqnarray}
\mathrm{KL}(p\|q) = \int_{z \in \mathcal Z} p(z) \log \frac{p(z)}{q(z)} \; dz .
\end{eqnarray}
When $q(z)$ is a Gaussian, selecting the mean and variance to be the same as those of $p(z)$ will minimize the divergence; this is sometimes known as ``moment matching.''  In particular, if $p(z)$ is a mixture of two Gaussians we have the following.

\begin{lemma}
The single Gaussian with mean $m$ and variance $v$ which minimizes the divergence from the mixture of two Gaussians $t_1 = (m_1,v_1,c_1)$ and $t_2=(m_2,v_2,c_2)$, with $c_1+c_2=1$, is given by: \label{lemma:mm}\end{lemma}
\begin{eqnarray}
m&=&c_1m_1+c_2m_2, \textrm{ and} \label{eqn:mean} \\
v&=&c_1 (m_1^2+v_1)+ c_2
(m_2^2+v_2) \nonumber \\ & &
-c_1^2m_1^2-2c_1c_2m_1m_2-c_2^2m_2^2. \label{eqn:variance}
\end{eqnarray}
The single Gaussian which satisfies the property of Lemma \ref{lemma:mm} is denoted as:
\begin{eqnarray}
t = \textrm{MM}(t_1, t_2),
\end{eqnarray}
where $t = (m,v,1)$, with $m$ and $v$ as given in $(\ref{eqn:mean})$ and $(\ref{eqn:variance})$.

While it is easy to find the mean and variance of the single Gaussian $q(z)$ which minimizes the divergence, the minimum divergence itself does not appear to have a closed form. Computing the divergence numerically is  complex.  Instead, the squared difference is computed.

\emph{Definition} The squared
difference $\textrm{SD}(p || q)$ between two distributions $p(z)$ and $q(z)$ with
 support $\mathcal Z$ is defined as:
\begin{eqnarray}
\textrm{SD}(p || q) &=& \int_{z \in \mathcal Z} (p(z) - q(z))^2 dz.
\end{eqnarray}

The squared difference is non-negative and zero if and only if $p=q$.   The squared difference is symmetric.

Computing the squared distance between a mixture of two Gaussians and a single Gaussian is  tractable. %(computing the $\mathcal L_1$ distance requires evaluation of error functions, and is not considered for complexity reasons).  
In Section \ref{sec:moregauss}, it will be convenient to have a distance measure, or penalty, associated with replacing a two-component Gaussian mixture by a single Gaussian.   Define the Gaussian quadratic loss $\textrm{GQL}(p)$ as the squared difference between a distribution $p$ and the Gaussian distribution with the same mean $m$ and variance $v$ as $p$:
\begin{equation}
\textrm{GQL}(p)=\textrm{SD}(p\|\,\mathcal{N}(m,v)).
\end{equation}

\begin{lemma}
For the distribution which is the mixture of two Gaussian functions $t_1=(m_1, v_1, c_1)$ and $t_2=(m_2, v_2, c_2)$ with $c_1 + c_2 = 1$, the Gaussian quadratic loss is given by:
\begin{eqnarray}
\lefteqn{\textrm{GQL}(t_1,t_2)=}  \nonumber \\
& & \frac{1}{2 \sqrt{\pi v}}
+ \frac{c_1^2}{2 \sqrt{\pi v_1}}
+ \frac{c_2^2}{2 \sqrt{\pi v_2}}  \nonumber \\
& & - \frac{2 c_1}{\sqrt{2 \pi (v+v_1)}} e^{ - \frac{ (m - m_1)^2}{2 (v+v_1)}}
- \frac{2 c_2}{\sqrt{2 \pi (v + v_2)}} e^{ - \frac{ (m-m_2)^2}{2 (v + v_2)}} \nonumber \\
& & + \frac{2 c_1 c_2}{\sqrt{2 \pi (v_1 + v_2)}} e^{ - \frac{(m_1- m_2)^2}{2(v_1+v_2)} } \label{eqn:sd},
\end{eqnarray}
where $m$ and $v$ are given in (\ref{eqn:mean}) and (\ref{eqn:variance}), respectively.
\label{lemma:gql}
\end{lemma}

When the mixture is not normalized, that is $c_1 + c_2 \neq 1$, the GQL is computed by first normalizing the two mixing coefficients to sum to one.

\newcommand{\mm}{^\mathsf{max}}

\subsection{Approximating $N$ Gaussians with $M$ Gaussians   \label{sec:moregauss} }

This section describes the Gaussian reduction algorithm which approximates a given mixture of $N$ Gaussian functions by a mixture of $M \leq N$ Gaussian functions. Specifically, the algorithm input $f(z)$ is the list of triples $t_i, i=1,\ldots,N$, and two stopping parameters $\theta$ and $M\mm$.  The algorithm output $g(z)$ is a mixture of $M$ Gaussian distributions, represented as a list of $M$ triples.  The algorithm is denoted as $g(z) = \mathrm{GMR}\big( f(z) \big)$.

\begin{figure}[t!]
\begin{center}
\includegraphics[width=8.3cm]{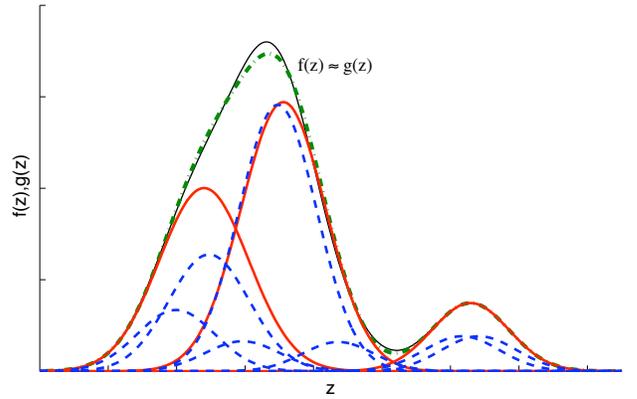}%
\end{center}
\caption{Seven Gaussian components (thick dashed line) approximated by three Gaussian components (thick solid line), using Gaussian mixture reduction algorithm with $\theta=0.01$.  The Gaussian mixture reduction output $g(z)$ (thick dash-dot line) is a good approximation of the input $f(z)$ (thin line). }
\label{fig:gaussian}
\end{figure}

The algorithm recursively combines Gaussians in a greedy pairwise fashion until two stopping conditions are fulfilled. At each recursion step, the pair of Gaussian distributions with the smallest GQL are eliminated, and are replaced by the single Gaussian with the same moments.  In this way, the GQL is a one-step error metric. The recursive procedure is halted when both of two conditions are satisfied. First, the GQL of all remaining pairs of Gaussians is greater than a combining limit $\theta$. Second, the number $M$ of Gaussian components in the current mixture is less than or equal to $M\mm$.  Note that even if the GQL of all remaining pairs of Gaussians is larger than $\theta$, the algorithms continues replacing until $M \leq M\mm$. In other words, we wish to avoid mixtures with many Gaussian components, even if this implies a large one-step GQL. On the other hand, the number of output Gaussians may be less than $M\mm$, if the one-step GQL is sufficiently low. An example of the input $f(z)$ and output $g(z)$ is shown in Fig.~\ref{fig:gaussian}.
The algorithm is as follows.

\noindent \emph{Gaussian Mixture Reduction Algorithm}
\begin{enumerate}
\item Input:
\begin{enumerate}
\item List $\mathcal L = \{t_1, t_2, \ldots, t_N\}$ of $N$ triples describing a Gaussian mixture.
\item Two stopping parameters, $\theta$ and $M\mm$.
\end{enumerate}
\item Initialize:
\begin{enumerate}
\item The current search list, $\mathcal C$, is the input list: $\mathcal C \leftarrow \mathcal L$.
\item The length of current list, $M^{\mathsf c}  \leftarrow N$.
\item The minimum GQL between all pairs of Gaussians, $\theta^{\mathsf c}$:
\begin{eqnarray*}
 \theta^{\mathsf c} \leftarrow \min_{t_i,t_j \in \mathcal C,i \neq j} \textrm{GQL}( t_i , t_j ).
 \end{eqnarray*}
\end{enumerate}
\item Greedy combining.  While $\theta^{\mathsf c} < \theta$ or $M^{\mathsf c} > M\mm$: \label{step:mmiter}
\begin{enumerate}
\item  Determine the pair of Gaussians $(t_i,t_j)$ with the smallest GQL:
\begin{eqnarray*}
(t_i,t_j) & \leftarrow& \arg \min_{t_i,t_j \in \mathcal C,i \neq j} \textrm{GQL}(t_i, t_j).
\end{eqnarray*}
\item Add the single Gaussian with the same moment as $t_i$ and $t_j$ to the list:\label{step:mm}
\begin{eqnarray*}
\mathcal C \leftarrow \mathcal C \cup \textrm{MM}(t_i, t_j).
\end{eqnarray*}
\item Delete $t_i$ and $t_j$ from list:  $\mathcal C \leftarrow \mathcal C \setminus \{t_i, t_j\}$.
\item Decrement the current list length:  $M^{\mathsf c}  \leftarrow M^{\mathsf c} - 1$.
\item Recalculate the minimum GQL:
\begin{eqnarray*}
 \theta^{\mathsf c} & \leftarrow& \min_{t_i,t_j \in \mathcal C,i \neq j} \textrm{GQL}( t_i ,  t_j ).
 \end{eqnarray*}
\end{enumerate}
\item Output: list of $M$ triples given by $\mathcal C$.
\end{enumerate}

This is a greedy algorithm, and no claims are made about its global optimality.  However, as demonstrated in the following sections, it has good performance when decoding LDLC lattices.

The primary complexity is proportional to $N^2$, due to the computation of the initial GQL between $N$ pairs at step 2-c.   Subsequent steps only compute the GQL between the new Gaussian and the remaining Gaussians.

\section{LDLC Lattices and Decoder}

\subsection{LDLC Lattices and Unconstrained Power System \label{sec:sys}}

An $n$-dimensional lattice $\Lambda$ is defined by an $n$-by-$n$ generator matrix $G$.   The lattice consists of the discrete set of points $\mathbf x = (x_1,x_2, \ldots, x_n)$ for which 
\begin{eqnarray}
\mathbf x &=& G \mathbf b, \label{eqn:lattice}
\end{eqnarray}
where $\mathbf b =(b_1,\ldots, b_n)$ is from the set of all possible integer vectors, $b_i \in \mathbb Z$.  Lattices are a linear subspace of the $n$-dimensional real space $\mathbb R^n$.

An LDLC lattice is a lattice with a non-singular generator matrix $G$, for which $H =  G^{-1}$ is sparse and random. It is convenient to assume that $H$ has been normalized so that $| \det(H)| = 1/ | \det(G)| = 1$. A Latin square LDLC code has an $H$ matrix with constant row and column weight $d$, where the non-zero coefficients in each row and each column have the values $h_1,h_2, \ldots, h_d$ with $h_1 \geq h_2 \geq \cdots \geq h_d > 0$.  The sign of each element of $H$ is randomly made negative with probability one-half \cite{Sommer-it08}.   In the numerical results section, Latin square LDLC codes with $h_1=1$ and $h_i = 1 / \sqrt{d}, i = 2,\ldots,d$, are used.

The unconstrained-power AWGN system, as was considered by Sommer et~al., is also used here.  An arbitrary element $\mathbf x$ of the lattice is transmitted over an AWGN channel.   Noise $w_i$ with variance $\sigma^2$ is added to each symbol.  The received symbols are $y_i = x_i + w_i$ for $ i=1,2,\ldots,n$.   A maximum-likelihood decoder estimates the transmitted lattice point $\widehat{\mathbf x} $ as
\begin{eqnarray}
\widehat{\mathbf x} = \arg \max_{\mathbf x \in \Lambda} Pr\big( \mathbf y | \mathbf x\big).
\end{eqnarray}

Since the transmit power is unbounded, the system is constrained by the lattice density, measured by the volume of the lattice's Voronoi region, $V_{\Lambda}=|\det(G)|$.   Somewhat analogous to the Shannon limit, there exist lattices such that the probability of error, $Pr\big(\mathbf x \neq \widehat{\mathbf x}\big)$ becomes arbitrarily small, if and only if,
\begin{eqnarray}
\sigma^2 &\leq & \frac{V_{\Lambda}^{2/n}}{2 \pi e}  ,\label{eqn:capacity}
\end{eqnarray}
provided the lattice dimension is allowed to grow without bound \cite{Tarokh-it99*2} \cite{Poltyrev-it94}.  This system is linear, so for finding error probabilities, it is sufficient to consider the transmission of the all-zeros lattice point.

\newcommand{\m}[3]{m_{#3} (#1_{#2}) }
\renewcommand{\v}[3]{v_{#3} (#1_{#2}) }
\renewcommand{\c}[3]{c_{#3} (#1_{#2}) }
\newcommand{\N}[2]{N(#1_{#2})}

\renewcommand{\r}{{\rho}}
\newcommand{\rt}{{\widetilde \rho}}
\newcommand{\q}{{\mu}}
\renewcommand{\a}{{\alpha}}
\newcommand{\at}{{\widetilde{\alpha}}}
\renewcommand{\b}{{\beta}}
\newcommand{\bt}{{\widetilde{\beta}}}

\subsection{LDLC Decoder Using Gaussian Mixtures \label{sec:ldlc}}

The LDLC belief-propagation decoding algorithm may be adapted to incorporate the Gaussian mixture algorithm.
% method of Section \ref{sec:moregauss}.   
The check node and variable node functions are decomposed into forward-backward recursions, and the Gaussian mixture reduction algorithm is applied after each recursion step.

The decoding algorithm may be presented on a bipartite graph with $n$ variable nodes and $n$ check nodes.  For Latin square LDLC codes, there are $nd$ variable-to-check messages $\q_k(z)$, and $nd$ check-to-variable messages $\r_k(z)$,  $k=1,2,\ldots, nd$.  Associated with variable node $i$ is the channel output~$y_i$, and channel message is $y_i(z) = \Nbb z {y_i} {\sigma^2} \label{eqn:channelmessage}$ ($y_i$ is distinct from $y_i(z)$).  The initial variable-to-check message for edge $k$ is $\q_k(z) = y_i(z)$, if edge $k$ is connected to variable node $i$.

\begin{figure}[t]
\begin{center}
\includegraphics[width=8.5cm]{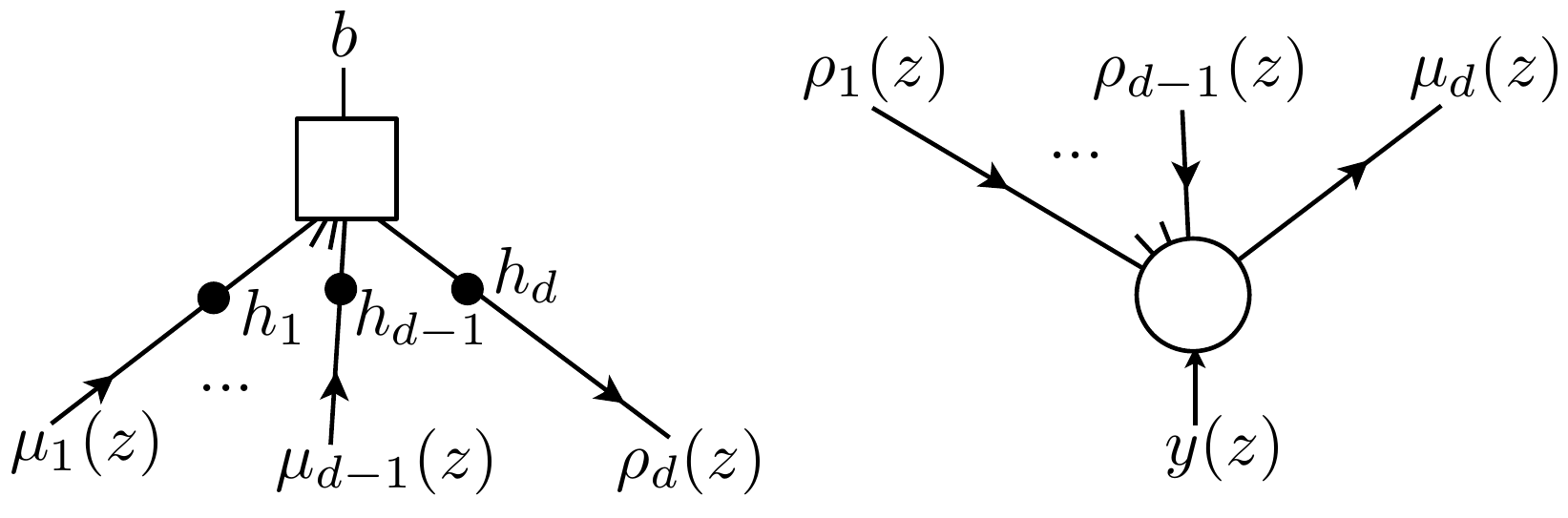}%
\\(a) \hspace{4cm} (b)
\end{center}
\caption{Messages for check node (a) and variable node (b) decoding functions. }
\label{fig:checkvarnode}
\end{figure}

\emph{Check node}  Corresponding to each edge $k=1,\ldots, d$ connected to a check node, the input messages are $\q_k(z)$, the output messages are $\r_k(z)$, and the edge labels from matrix $H$ are $h_k$, as shown in Fig. \ref{fig:checkvarnode}-(a).  The check node output along some edge is the convolution of the incoming messages on all other edges, followed by a shift-and-repeat (periodic extension) operation.  For example, for edge $d$, the convolution before shift-and-repeat $\rt_d(z)$ is:
\begin{eqnarray}
\rt_d(z) &=& \q_1\big( \frac{z}{h_1} \big) * \cdots * \q_{d-1}\big( \frac{z}{h_{d-1} } \big) \label{eqn:convolution}
\end{eqnarray}
An unknown integer $b$ is from the set $\mathcal B$; in the general case $\mathcal B$ is the set of all integers.   The shift-and-repeated message $\r_k(z)$ along edge $k$ is :
\begin{eqnarray}
\r_k(z) &=&  \sum_{b \in \mathcal B} {\rt}_k(b - h_k z),  \label{eqn:gmdcheck}
\end{eqnarray}
which has a period of $1/| h_k| $ if $\mathcal B$ is the set of integers.

\renewcommand{\u}{{\a}}
\renewcommand{\v}{{\q}}

The check node convolution (\ref{eqn:convolution}) can be implemented by a forward-backward recursion as follows. First, denote a single convolution as:
\begin{eqnarray}
\textrm C\Big( \u(z), \v\big(\frac z h \big) \Big) &=& \u(z) * \v \big(\frac{z}{h}\big).
\end{eqnarray}
Then, the forward recursion is initialized with $\alpha_0 = \big\{ (0,0,1) \big\}$, in triples representation. The forward recursion step is convolution followed by Gaussian mixture reduction,
\begin{eqnarray}
\at_k(z) &=& \textrm C \Big( \a_{k-1}(z) , \q_k\big( \frac z {h_k}\big) \Big), \label{eqn:checkforward} \\
\a_k(z) &=& \textrm{GMR}\big(\at_k(z) \big),
\end{eqnarray}
for $k=1,\ldots,d-1$.  Backward messages $\b_k(z)$ and $\bt_k(z)$ are found in an analogous fashion, initialized with  $\beta_d=\{ (0,0,1) \}$, with the recursion,
\begin{eqnarray}
\bt_k(z) &=& \textrm C \big( \b_{k+1}(z) , \q_{k+1}(z) \big), \label{eqn:checkforward} \\
\b_k(z) &=& \textrm{GMR}\big(\bt_k(z) \big),
\end{eqnarray}
for $k=d-1,\ldots,2,1$.

The convolution, $\textrm C\big( \a(z), \q(z / h) \big)$, in triples representation $\big\{ ( m_\ell, v_\ell, c_\ell ) \big\}$, can be computed as follows.   Let $\u(z)$ be a mixture of $I$ Gaussians, and let $\v(z)$ be a mixture of $J$ Gaussians, represented as:
\begin{eqnarray}
\Big\{ \big(m^\u_i, v^\u_i, c^\u_i\big) \Big\}_{i=1}^I
& \textrm{and} &
\Big\{ \big(m^\v_j, v^\v_j, c^\v_j \big) \Big\}_{j=1}^J,
\end{eqnarray}
respectively.  Then, the convolution is a mixture of $I \cdot J$ Gaussians, given by:
\begin{eqnarray}
m_{\ell} &=& m^\u_i + {h} \cdot {m^\v_j}, \\
v_{\ell}  &=& v^\u_i + {h^2}\cdot {v^\v_j}, \\
c_{\ell} &=& c^\u_i \cdot c^\v_j,
\end{eqnarray}
where each component $i \in \{1,\ldots,I\}$ of $\u(z)$ and each component $j \in \{1,\ldots,J\}$ of $\v(z)$ is pair-wise convolved to produce an output component $\ell \in \{1,\ldots,I \cdot J\}$.

%\at_k(z) &=& \a_{k-1}(z) * \q_k(\frac{z}{h_k}), \label{eqn:checkforward} \\

The forward-backward recursions are completed by computing the message $\rt_k(z)$ as:
\begin{eqnarray}
\rt_k(z) &=& \textrm{GMR} \big( \a_{k-1}(z) * \b_k(z) \big).
\end{eqnarray}
The check node output $\r_k(z)$ is found by applying the shift-and-repeat operation $(\ref{eqn:gmdcheck})$, which increases the number of Gaussians by a factor of  $| \mathcal B |$,  where $| \mathcal B|$ is the cardinality of the set $\mathcal B$.

\renewcommand{\u}{{\a}}
\renewcommand{\v}{{\r}}

\emph{Variable Node}   Corresponding to each edge $k=1,\ldots, d$ connected to a variable node, the inputs are messages $\r_k(z)$ and the outputs are messages $\q_k(z)$; the channel message is $y(z)$, as shown in Fig. \ref{fig:checkvarnode}-(b).  The variable node output along some edge is the product of the incoming messages on all other edges, including the channel message.  For example, the output for edge $d$, $\q_d(z)$, is:
\begin{eqnarray}
\q_d(z) &=& y(z) \cdot \r_1(z) \cdot \ldots \cdot \r_{d-1}(z) \label{eqn:varmult}
\end{eqnarray}

\begin{figure*}[t]
\begin{center}
\includegraphics[width=14.3cm]{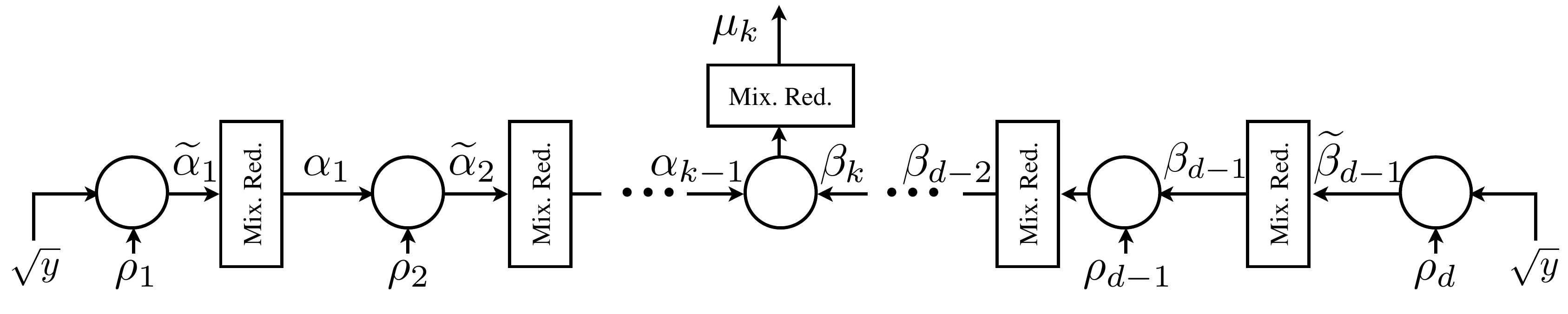}%
\end{center}
\caption{Forward-backward variable node decoding, with Gaussian mixture reduction applied after each step. }
\label{fig:fbmr}
\end{figure*}

The variable node multiplication (\ref{eqn:varmult}) can be implemented by a forward-backward recursion as follows, and is illustrated in Fig.~\ref{fig:fbmr}.  First, denote a single multiplication as:
\begin{eqnarray}
\textrm V\big( \u(z), \v( z ) \big) &=& \u(z) \cdot \v (z).
\end{eqnarray}
Then, the forward recursion is initialized with $a_0(z) = \sqrt{y(z)}$. The square root of a Gaussian distribution is another Gaussian distribution with the same mean and twice the variance, so in triples representation the initialization is $\{ ( y, 2 \sigma^2,1) \}$.   Then the forward recursion for $k=1,2\ldots,d$ is,
\begin{eqnarray}
\at_k(z) &=& \textrm V \big( \a_{k-1}(z) ,  \r_k(z) \big), \textrm{ and} \label{eqn:varforward}\\
\a_k(z) &=& \textrm{GMR}(\at_k(z)).
\end{eqnarray}
The backward recursion similarly computes messages $\b_k$, and is initialized with $\b_d(z) = \sqrt{y(z)}$, and the recursion is:
\begin{eqnarray}
\bt_k(z) &=& \textrm{V} \big( \b_{k-1}(z) , \r_k(z) \big) \textrm{ and} \label{eqn:varforward} \\
\b_k(z) &=& \textrm{GMR}(\bt_k(z)).
\end{eqnarray}

The product $\textrm V\big( \a(z), \r(z) \big)$, in triples representation $\big\{ ( m_\ell, v_\ell, c_\ell ) \big\}$, can be computed as follows.   Let $\u(z)$ be a mixture of $I$ Gaussians, and let $\v(z)$ be a mixture of $J$ Gaussians, represented as:
\begin{eqnarray}
\Big\{ \big(m^\u_i, v^\u_i, c^\u_i\big) \Big\}_{i=1}^I
& \textrm{and} &
\Big\{ \big(m^\v_j, v^\v_j, c^\v_j \big) \Big\}_{j=1}^J,
\end{eqnarray}
respectively.    Then, the product is a mixture of $I \cdot J$ Gaussians, given by:
\begin{eqnarray}
\frac{1}{v_{\ell}} &=& \frac{1}{v^\u_i} + \frac{1}{v^{\v}_j}, \\
\frac{m_{\ell}}{v_{\ell}} &=& \frac{m^{\u}_i}{v^\u_i} + \frac{m^{\v}_j}{v^{\v}_j}, \\
c_{\ell} &=& \sqrt{  \frac{1} {2 \pi (v^{\u}_i + v^{\v}_j)  } } \exp \Big( \frac 1 2 \frac{ (m_i^{\u} - m_j^{\v} )^2 }{  v_i^{\u} + v_j^{\v} } \Big),
\end{eqnarray}
where each component $i \in \{1,\ldots,I\}$ of $\u(z)$ and each component $j \in \{1,\ldots,J\}$ of $\v(z)$ is pair-wise multiplied to produce an output component $\ell \in \{1,\ldots,I \cdot J\}$.

The forward-backward recursions are completed by computing the output message $\q_k(z)$ as:
\begin{eqnarray}
\q_k(z) &=& \textrm{GMR} \big( \a_{k-1}(z) \cdot \b_k(z) \big) .
\end{eqnarray}

The incoming messages $\r_k(z)$ have a large number of Gaussian components, created by the shift-and-repeat operation.   However, by initializing the recursion with a single Gaussian, many components away from this Gaussian have near-zero mixing coefficients, and are eliminated or combined by the Gaussian mixture reduction algorithm.    This effect persists as the forward-backward recursions proceed, and in both directions.     This would not occur in the backward recursion using the more naive initialization of $\a_0(z) = y(z)$ and $\b_d(z) =1$. Note that the output message $\r_k(z)$ is the product $\a_{k-1}(z) \cdot \b_k(z)$; since both $\a_{k-1}(z)$ and $ \b_k(z)$ include the factor $\sqrt{y(z)}$, $\r_k(z)$ correctly includes the factor $y(z)$.

\emph{Hard Decisions}  Hard decisions  on the integers, $\widehat{\mathbf b}$, may be found by first making  decisions $\widetilde{\mathbf x} = \left( \widetilde x_1, \ldots, \widetilde x_n\right)$, as,
\begin{eqnarray}
\widetilde x_i &=& \arg \max_z y(z) \prod_{k=1}^d \rho_k(z), \label{eqn:hardoutput}
\end{eqnarray}
where $y(z)$ and $\rho_k(z)$ are the channel message and input messages, respectively, for variable node $i$.   Practically, this product may be found as $\a_d(z) \sqrt{y(z)}$.    Then, $\widehat{\mathbf b}$ may be found by rounding each element of the vector $H \widetilde{\mathbf x}$ to the nearest integer.  A ``word error'' occurs if, after a sufficiently large number of iterations, $\mathbf b \neq \widehat{\mathbf b}$.   Note that $\widetilde{\mathbf x}$ is not necessarily a lattice point.

\section{Numerical Results \label{sec:numerical}}

\subsection{Finite-dimensional Lattices \label{sec:finite} }

\begin{figure}[t]
\begin{center}
\includegraphics[width=8.5cm]{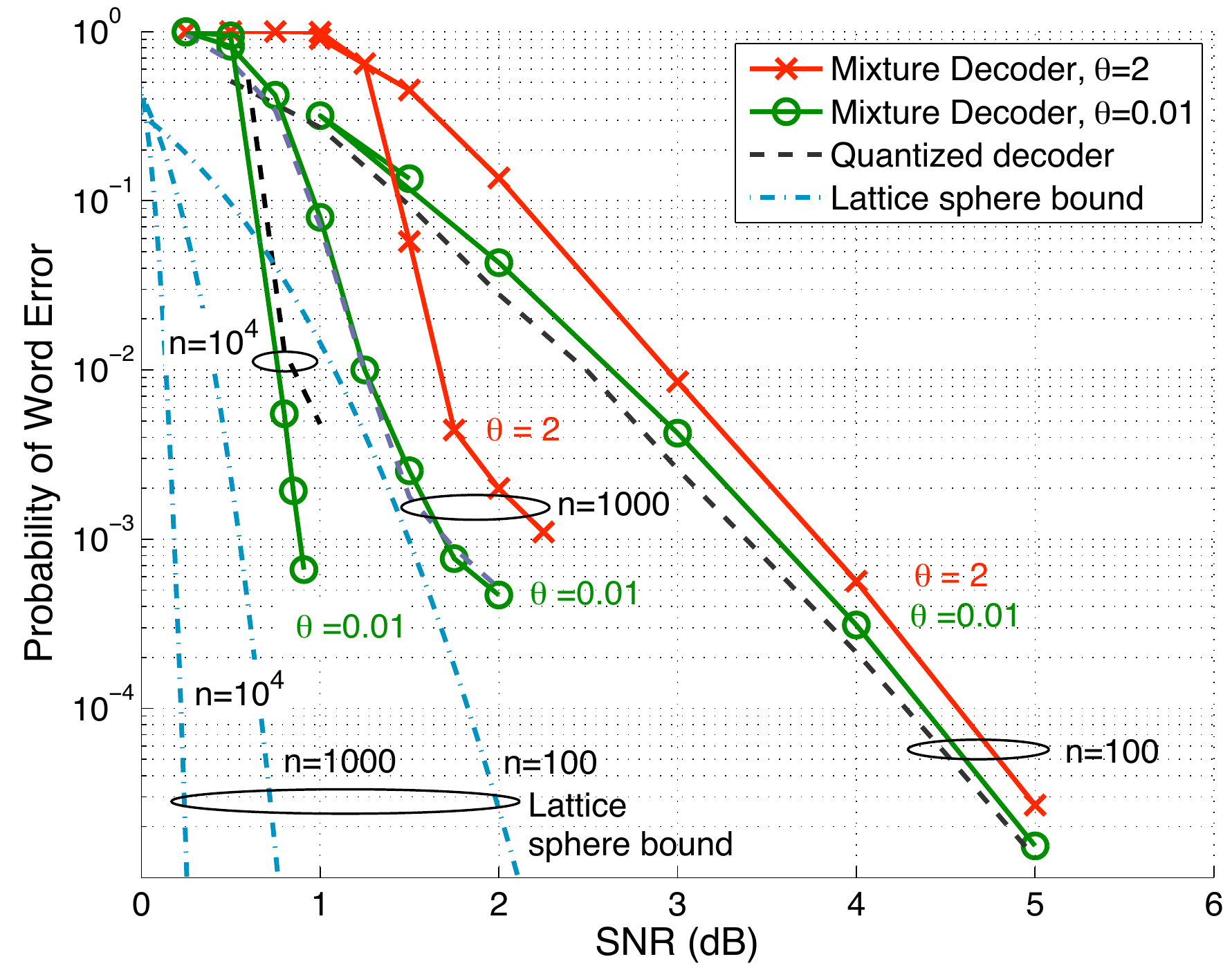}%
\end{center}
\caption{Probability of word error for Gaussian mixture decoder and quantized-message decoder.}
\label{fig:gmsim}
\end{figure}

For finite-dimensional lattices, the Gaussian mixture decoder has error rates similar to the quantized decoder, as shown in Fig.~\ref{fig:gmsim}.   For a dimension 100 lattice with $d=5$, the performance loss was no greater than 0.2 dB.   For a dimension 1000 and 10000 lattice both with $d=7$, there was no discernible performance loss at high signal-to-noise ratios when $\theta = 0.01$.   Fig. \ref{fig:gmsim} shows the word error probability $Pr(\mathbf b \neq \widehat{\mathbf b})$, rather than error rates for the symbols $b_i$.    The SNR is  $V_{\Lambda}^{2/n} / 2 \pi e \sigma^2$, and for all lattices $H$ is normalized such that $V_{\Lambda} = 1$.   When the channel noise variance $\sigma^2$ is equal to the capacity, that is, it satisfies eqn (\ref{eqn:capacity}) with equality, the SNR is 0 dB.  

Instabilities in the decoding algorithm can occur, but these may be avoided by setting a minimum message variance.  As iterations progress, Gaussian variances will decrease towards narrow peaks.  But due to quantization effects, two narrow peaks may not align and the Gaussian mixture reduction algorithm will fail to combine them.  This was avoided by setting a minimum variance.  A message variance $v$ is replaced by $\max ( v, \gamma)$, where $\gamma$ is a constant.   Values in the range $\gamma = 10^{-3}$ to $10^{-4}$ were used.  The quantized-message decoder uses a distinct method to avoid instabilities \cite[p.~1573]{Sommer-it08}.

Tarokh, Vardy and Zeger gave a lattice sphere bound, which is a lower bound on the probability of word error for the unconstrained-power system \cite[Theorem 2.2]{Tarokh-it99*2}.   This bound was shown to be reasonably tight for $n=16$, and is expected to further improve for increasing dimension.   This lattice sphere bound is also shown in Fig. \ref{fig:gmsim}.  For $n=100$ LDLC lattices, there is a substantial gap of about 2.5 dB, suggesting that there is room for improvement in LDLC lattice design, belief propagation decoding, or both.   However, this gap decreases as the lattice dimension increases.

\subsection{Complexity-Performance Tradeoff   \label{sec:complexity} }

The complexity-performance tradeoff for decoding is obtained through the combining limit $\theta$.  Decreasing $\theta$ increases the complexity and improves performance. If $\theta$ is high, mixtures are allowed to  contain many components which accurately represent the message and fewer decoding errors are made, however, the computational complexity is then substantial.   On the other hand, lower $\theta$ will lead to decoders with lower complexity but with higher error probability.   The maximum number of allowed Gaussians $M\mm$ was set to a large number, for example $M\mm=1000$, without an obvious increase in complexity.

Noise thresholds are used to characterize the performance of LDLC lattices.   The noise threshold is the lowest SNR for which density evolution of an asymptotically large dimension lattice converges.  For binary low-density parity-check codes on the AWGN channel, density evolution can be used, because the decoder messages are scalars, and the density (or distribution), can be discretized \cite{Richardson-it01}.   However for non-binary low-density parity-check codes, such as those constructed over finite fields, the messages are vectors.  True density evolution would use a discretized joint density over a vector, which is impractical.   Nonetheless, although less inefficient, noise thresholds for non-binary low-density parity-check codes may be found using Monte Carlo density evolution \cite{Davey-99}.    

The situation for the Gaussian-mixture LDLC decoder is similar to non-binary low-density parity-check codes, because the decoder messages consist of multiple means, variances and mixing coefficients.   Accordingly, LDLC lattice noise thresholds are found by Monte Carlo density evolution, as follows.  At each half iteration samples for $N_{\mathrm s} = 10^6$ nodes were randomly drawn from an input pool, and then placed in an output pool.  The pool has $d$ types of messages to distinguish the edges with distinct coefficients $h_1$ to $h_d$. The output pool becomes the input pool for the next half iteration, and this procedure repeats until convergence was detected.   Each variable node produces an output $\widetilde x_i$ given by (\ref{eqn:hardoutput}), and the mean-squared error is computed as $\sum_i^{N_{\mathsf s}} x_i^2 / {N_{\mathsf s}}$, since the all-zeros lattice point is assumed.   Convergence is declared when the mean-squared error approaches zero.   Interest is restricted to LDLC lattices with $h_1 = 1$ and $h_2=\cdots = h_d = 1 / \sqrt{d}$.  As such, power is suitably normalized, since such lattices have $1 / | \det H | = V_\Lambda \rightarrow 1$ as the dimension becomes large.   

The complexity of the decoding algorithm is proportional to $M^4$, where $M$ is the number of Gaussians in the input mixtures.   In the forward-backward recursions, the convolution or product of two inputs with $M$ Gaussians each produces a mixture with $M^2$ Gaussians, for example eqns. (\ref{eqn:checkforward}), (\ref{eqn:varforward}).   This mixture of $M^2$ Gaussians is the input to the Gaussian mixture reduction algorithm.   Since the complexity of Gaussian mixture reduction algorithm is proportional to the square of the number of inputs, the net complexity is $M^4$.   However, $M$ is usually much less than $M\mm$, and the distribution of $M$ indirectly depends on the combining limit $\theta$, channel noise $\sigma^2$, the iteration number and the code design.

\begin{figure}[t]
\begin{center}
\includegraphics[width=8.5cm]{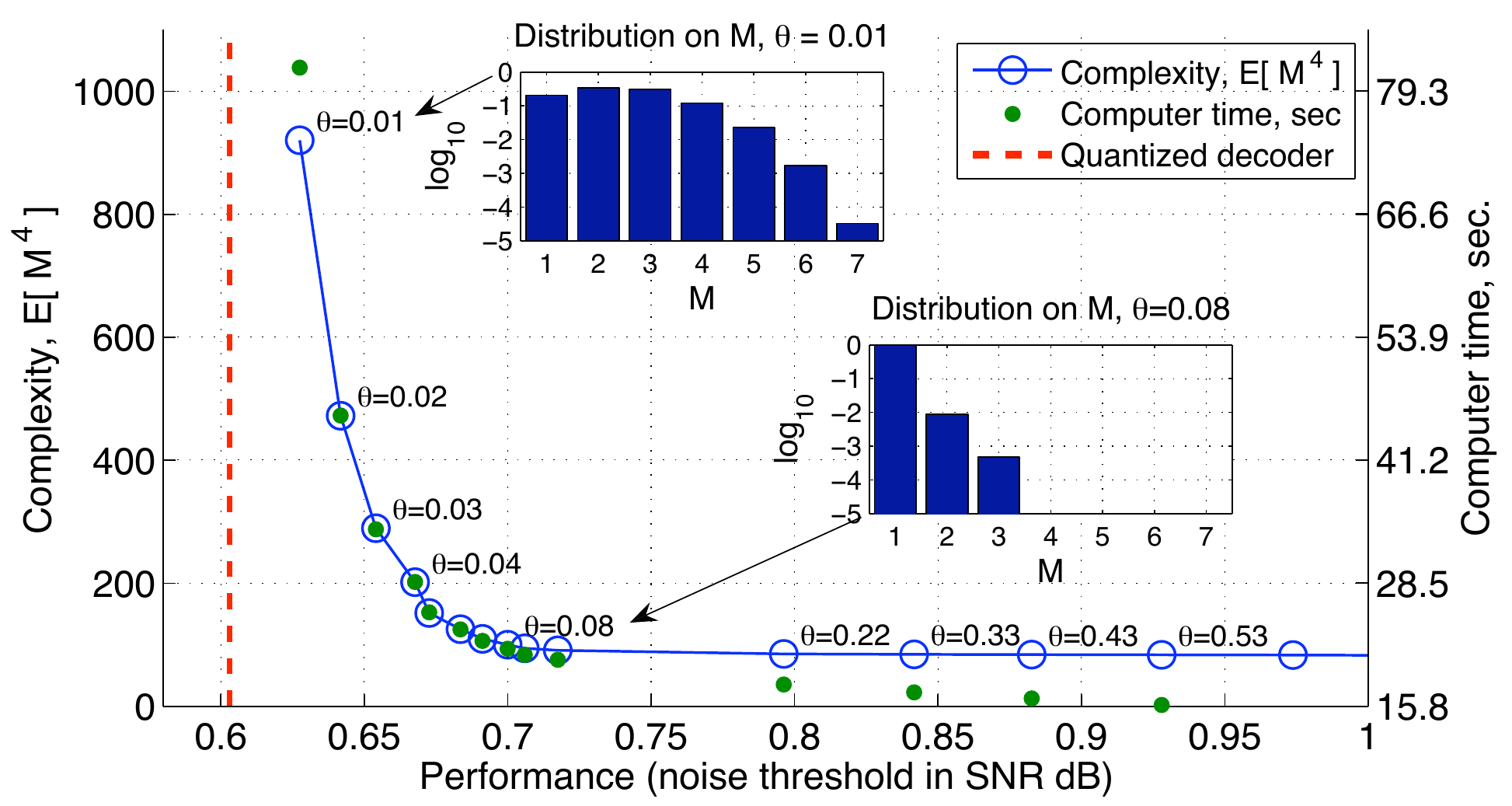}%
\end{center}
\caption{Performance-complexity tradeoff, parameterized by Gaussian combining parameter $\theta$; $d=7, M\mm = 1000, \mathcal B = \{-1,0,+1\}$.  Performance measure is noise threshold, found by Monte Carlo density evolution.  The complexity $E[M^4]$ (left-hand axis) is a reasonable predictor of computer time  (per iteration, right-hand axis); the two y-axes have different zeros due to computer overhead.  The inset graphs show representative distributions on $M$ for the message $\rt(z)$ associated with edge $h_1$, on iteration 5, for two values of $\theta = 0.01, 0.08$.}
\label{fig:mdist}
\end{figure}

Accordingly, the expected value of $M^4, \mathrm E [M^4]$, is used to characterize complexity.  During Monte Carlo density evolution, the distribution on the number of Gaussians in the message was collected (see Fig.~\ref{fig:mdist} insets), and used to find $\mathrm E[M^4]$.    Many messages $\q_k(z)$ contained a few Gaussian components. Likewise, the number of Gaussian components of $\r_k(z)$ was a small integral multiple of $|\mathcal B|$.  Because these latter messages contain more Gaussians, complexity is dominated by the variable node.

The performance-complexity tradeoff is illustrated in Fig.~\ref{fig:mdist}.  For small values of the combining limit $\theta = 0.01$, Gaussian-mixture decoding has a noise threshold within 0.03 dB of the quantized decoder.   On the other hand, choosing $\theta= 0.1$ increases the noise threshold loss to 0.1 dB, while substantially reducing the complexity; further increase in $\theta$ results in more performance loss with little complexity improvement.  A noise threshold of 0.6 dB was found for the quantized-message decoder, which corresponds to the symbol-error rate cliff of a dimension $10^5$ lattice simulated by Sommer et~al.

The complexity of the quantized algorithm of Sommer et~al. is dominated by a discrete Fourier transform at the check node, and thus it is difficult to make complexity comparisons.   However, the memory required for the Gaussian mixture-based LDLC decoder is significantly superior, storing $3M$ values (for the mean, variance and mixing coefficient), for each message.  On the other hand, the quantized algorithm used 1024 quantization points for each message.

\section{Discussion \label{sec:discussion}}

Codes based upon lattices can achieve the capacity of the power-constrained AWGN channel, as has been shown by Loeliger \cite{Loeliger-it97}, Erez and Zamir \cite{Erez-it04}, and others.  In fact, high-dimension lattices are good in a number of information-theoretic contexts \cite{Erez-it05*4}.   However, for most lattices other than LDLC lattices, the decoder complexity is worse than linear in the dimension, and so decoding high-dimension lattices to approach capacity is infeasible.

Sommer et~al. demonstrated that decoding high-dimension lattices is possible by using a linear-complexity belief propagation algorithm.   Moreover, the error-rate performance of LDLC lattices comes with 0.6 dB of the capacity of the unconstrained power channel.   The unconstrained-power channel is of significant theoretical interest \cite{Poltyrev-it94}, but further study of LDLC lattices is needed before they can be applied to more practical problems such as the power-constrained AWGN channel.

The quantized-message decoder used to demonstrate the capacity-approaching performance ignores the underlying Gaussian nature of the messages.     This paper demonstrated that it is possible to perform belief-propagation decoding of LDLC lattices using a mixture of Gaussian functions as the decoder message.    The key part was the Gaussian mixture reduction algorithm, which was proposed to approximate a mixture of a large number of Gaussians with a smaller number of such Gaussians.   Although it is an approximation, numerical studies showed this method to have little or no performance loss.  The representation of messages as a Gaussian mixture uses far fewer parameters than the quantized-message decoder, which improves the efficiency of the decoder.  In addition, belief-propagation decoding using a Gaussian mixture characterized by a small number of parameters should advance further study of LDLC lattices.

%\bibliographystyle{ieeetr}
%\bibliographystyle{plain}
%\bibliography{bibtex/abbrev,bibtex/abbrev_brian,bibtex/starbib,bibtex/brian}

\begin{thebibliography}{1}

\bibitem{Sommer-it08}
N.~Sommer, M.~Feder, and O.~Shalvi, ``Low-density lattice codes,'' {\em IEEE
  Transactions on Information Theory}, vol.~54, pp.~1561--1585, April 2008.

\bibitem{Scott-technometrics01}
D.~W. Scott and W.~F. Szewczyk, ``From kernels to mixtures,'' {\em
  Technometrics}, vol.~43, pp.~323--335, August 2001.

\bibitem{Sarvotham-TR-Rice-2006}
S.~Sarvotham, D.~Baron, and R.~G. Baraniuk, ``Compressed sensing reconstruction
  via belief propagation,'' Tech. Rep. ECE-06-01, Department of Electrical and
  Computer Engineering, Rice University, July 2006.

\bibitem{Goldberger-pami08}
J.~Goldberger, H.~K. Greenspan, and J.~Dreyfuss, ``Simplifying mixture models
  using the unscented transform,'' {\em IEEE Transactions on Pattern Analysis and Machine
  Intelligence}, vol.~30, pp.~1496--1502, Aug. 2008.

\bibitem{Tarokh-it99*2}
V.~Tarokh, A.~Vardy, and K.~Zeger, ``Universal bound on the performance of
  lattice codes,'' {\em IEEE Transactions on Information Theory}, vol.~45,
  pp.~670--681, March 1999.

\bibitem{Poltyrev-it94}
G.~Poltyrev, ``On coding without restrictions for the {AWGN} channel,'' {\em
  IEEE Transactions on Information Theory}, vol.~40, pp.~409--417, March 1994.

\bibitem{Richardson-it01}
T.~J. Richardson and R.~L. Urbanke, ``The capacity of low-density parity check
  codes under message-passing decoding,'' {\em IEEE Transactions on Information
  Theory}, vol.~47, pp.~599--618, February 2001.

\bibitem{Davey-99}
M.~C. Davey, {\em Error-correction using low-density parity-check codes}.
\newblock PhD thesis, University of Cambridge, 1999.

\bibitem{Loeliger-it97}
H.-A. Loeliger, ``Averaging bounds for lattices and linear codes,'' {\em IEEE
  Transactions on Information Theory}, vol.~43, pp.~1767--1773, November 1997.

\bibitem{Erez-it04}
U.~Erez and R.~Zamir, ``Achieving $\frac{1}{2}\log (1+ \textrm{SNR})$ on the
  {AWGN} channel with lattice encoding and decoding,'' {\em IEEE Transactions
  on Information Theory}, vol.~50, pp.~2293--2314, October 2004.

\bibitem{Erez-it05*4}
U.~Erez, S.~Litsyn, and R.~Zamir, ``Lattices which are good for (almost)
  everything,'' {\em IEEE Transactions on Information Theory}, vol.~51,
  pp.~3401--3416, October 2005.

%\bibitem{Cover-1991}
%T.~M. Cover and J.~A. Thomas, {\em Elements of Information Theory}.
%\newblock Wiley, 1991.

\end{thebibliography}

\end{document}